\documentclass[journal]{IEEEtran}
\usepackage{cite}
\usepackage{amsmath,amssymb,amsfonts}
\usepackage{graphicx}
\usepackage{hyperref}
\begin{document}

\title{A public key infrastructure for compliant spectrum use by wireless devices}

\author{\IEEEauthorblockN{Stefan Ruehrup}

\IEEEauthorblockA{
\textit{ITS Services} \\
\textit{ASFINAG Maut Service GmbH}\\
Vienna, Austria } \\
}

\maketitle

\begin{abstract}
A multitude of wireless devices for radio local area networks (RLAN) use frequency bands that are shared with other radio services or applications. European market surveillance activities on 5 GHz RLAN indicate a growing number of non-compliant devices that do not fulfill all radio regulation requirements and may cause interference. 

This paper describes a concept and method for mitigating the use of non-compliant equipment based on a public key infrastructure. If every RLAN device holds a security certificate, which is revoked when non-compliance is found, then the connection establishment between devices can be made dependent on compliance. The aim is that legitimate equipment rejects connections to non-compliant devices, so that non-compliant devices will become isolated. 
\end{abstract}


\section{Introduction}

Wireless local area networks (WLANs) are one of the most popular means to provide Internet connectivity to mobile and portable devices. WLAN devices use frequency bands assigned to Wireless Access Systems and Radio Local Area Networks (WAS/RLAN) \cite{ecc-dec-04-08}. These frequency bands are shared with other radio services or applications, and certain mitigation techniques are required by regulation, such as Dynamic Frequency Selection (DFS) to protect radars. 

European market surveillance activities conducted in 2018 \cite{adcored2019} indicated a growing number of non-compliant 5 GHz RLAN devices that do not fulfill all requirements of the radio equipment directive. If non-compliant equipment is brought to the market and used, it may  cause harmful interference to other radio services and applications. 

In a concrete case of interference, the search for its source might become as difficult as finding a needle in a haystack. First, because there are lots of devices, since WLANs are present in most offices and households. Second, these devices are not alway turned on, but rather used sporadically. Therefore it is difficult to detect and locate a single non-compliant device. This problem serves as motivation for this paper. We will abstract from the concrete example and use in the following only the terms ``compliant'' and ``non-compliant'' devices for radio equipment that does not fulfill the requirements of the radio regulation (which is different from compliance to communication standards).  

In order to mitigate this problem, the main idea and method described in this paper is to  establish a ``trusted'' environment using digital certificates, in which all compliant equipment operates. Non-compliant equipment, once identified, is not part of or removed from this environment an becomes isolated. 

\begin{figure}[htbp]
\centerline{\includegraphics[width=\linewidth]{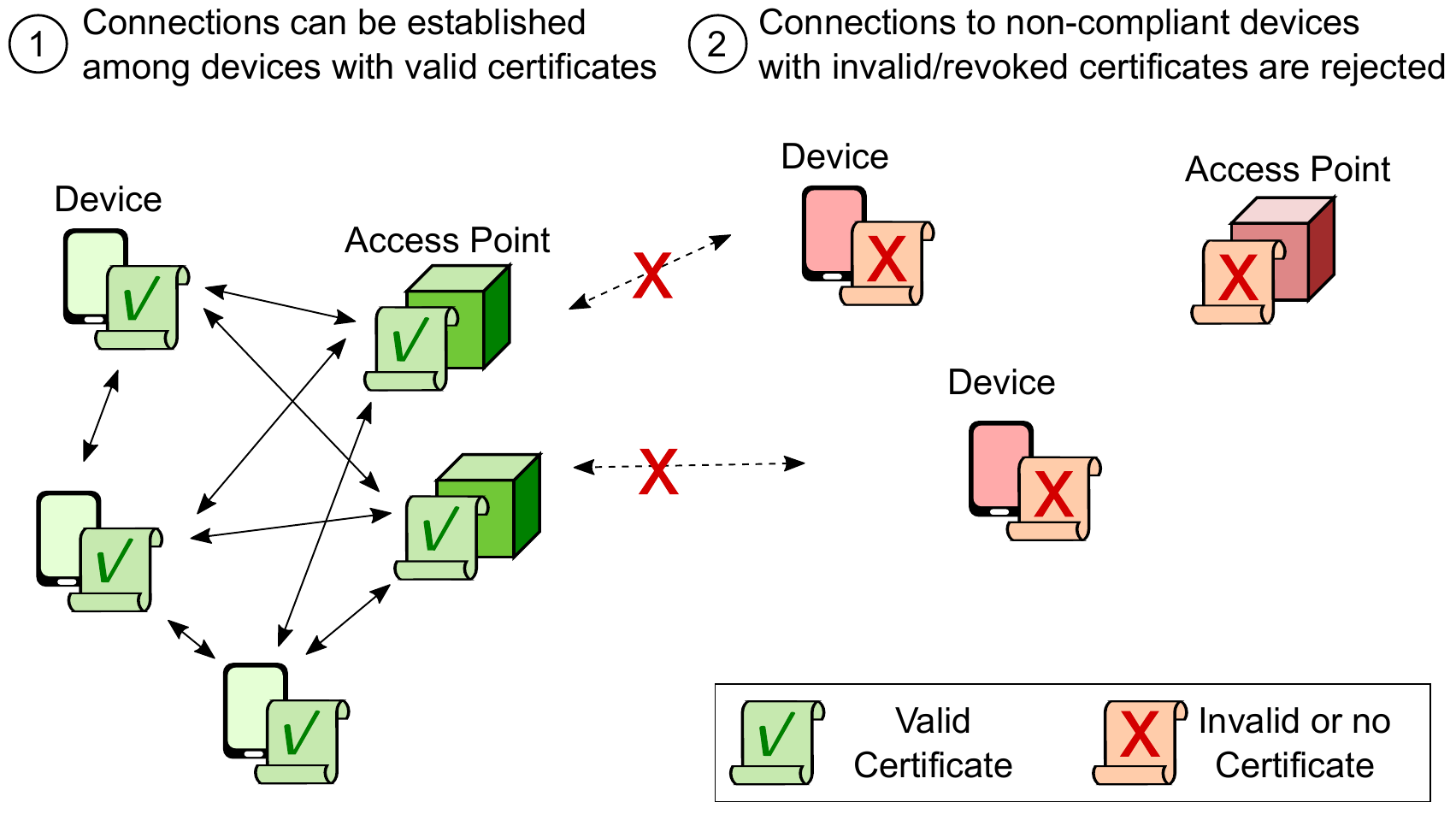}}
\caption{Isolation of non-compliant devices: Compliant devices with valid certificates communicate to each other, but reject connections to devices without valid certificate.}
\label{fig:isolation}
\end{figure}

Such environment can be built based on public key infrastructures (PKI). A PKI is a hierarchy of digital certificates, where each certificate consist of a public key and a signature by a higher level certificate authority. Using  for verifying digital signatures -- a valid signature indicates a valid certificate. Assume that wireless devices (access points, tablets etc.) hold such certificates and use them create digital signatures on data exchanged during or after connection establishment. Then, a device can verify whether the signature and thus the certificate of its counterpart is valid and maintain the connection, otherwise it drops the connection. Once a device model is found to be non-compliant, the certificate for this model is revoked, and this device model cannot connect to other legitimate devices any longer. This way, the legitimate devices reject connections and non-compliant devices become isolated (see Fig.~\ref{fig:isolation}). 

This paper describes the overall concept and architecture of using PKIs and trust lists, which link PKIs together. One option for implementation is the following: A PKI issuing certificates to devices can be maintained by a vendor while a trust list of root certificates links the PKIs together. The overall trust list could be maintained by an industry association or a government entity.

This paper is structured as follows:
In Section~\ref{sec:related} we review related work. 
In Section~\ref{sec:concept} we describe the overall architecture. 
In Section~\ref{sec:discussion} we discuss implementation aspects. 
Section~\ref{sec:conclusion} concludes the paper.

\section{Related Work} \label{sec:related}

Public key infrastructures (PKIs) and trust lists are established mechanisms to setup and maintain a trust environment with respect to cybersecurity, see \cite{mazaher2003} for a survey. They are used in the Internet to established secure connections, and Internet browsers usually have the capability to import certificates and validate signatures. PKIs also provide the basis for digital signatures. There are standards and recommendations for protocols, data formats and crypto-algorithms. The most common standard for certificates is the ITU recommendation X.509 \cite{x509}. Formats for trust lists and certificate revocation lists (CRLs) are specified, e.g., in \cite{rfc5914} and \cite{rfc5280}.

Trust lists are also used in the context of digital signatures. As an example, status information of certification service providers in the EU are maintained in trusted lists defined in Commission Decision 2009/767/EC and its implementing act \cite{C(2013)6543}, based on the technical specification in ETSI TS 119 612 \cite{etsi-ts-119-612}. These trusted lists are maintained by the Member States.

\begin{figure}[htbp]
\centerline{\includegraphics[width=\linewidth]{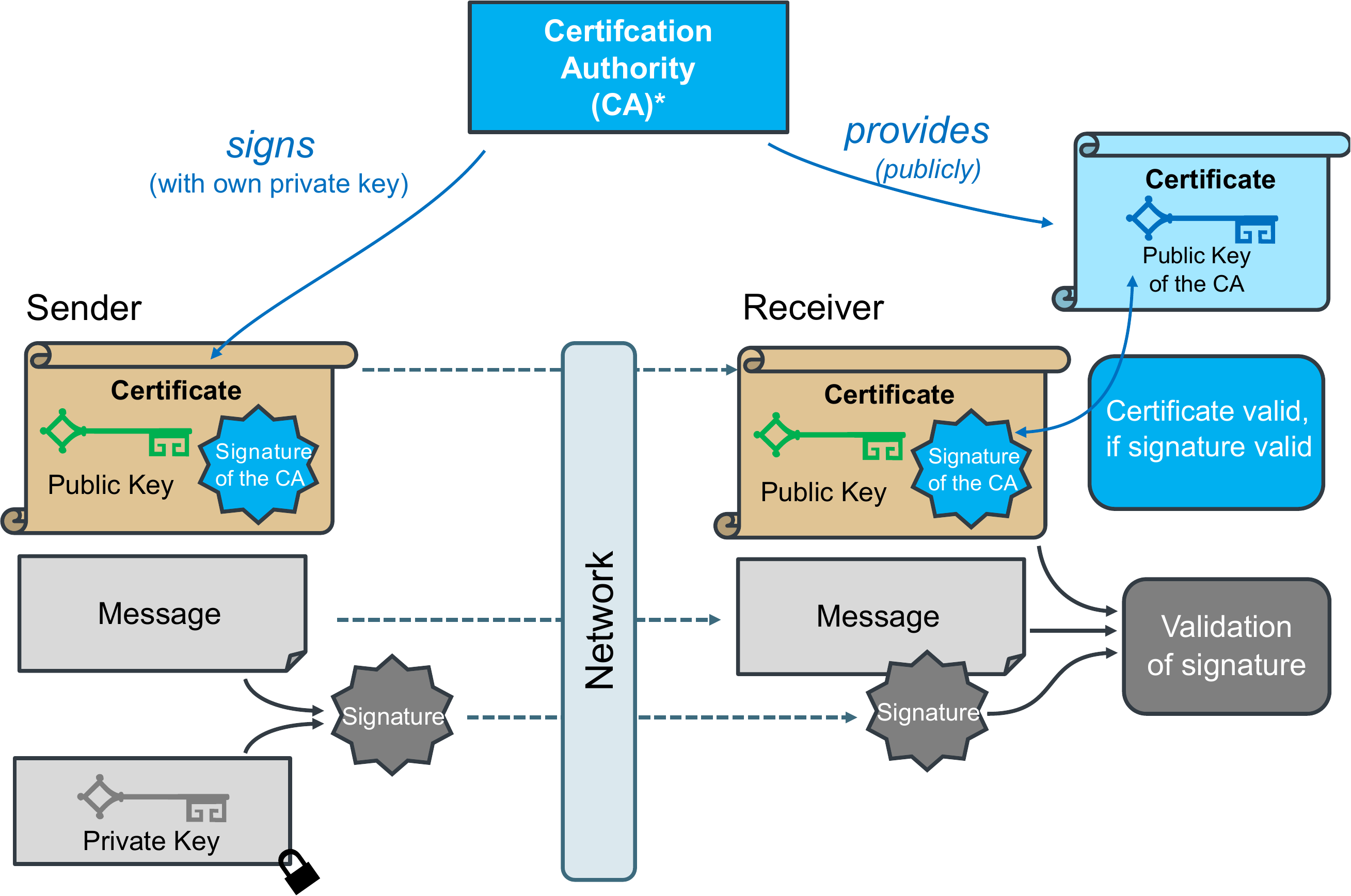}}
\caption{Validation of a signature: The receiver validates the signed message by decrypting the signature with the help of the public key of the sender. The public key of the sender is signed by the CA and can be validated if it is part of the trust list. }
\label{fig:cert-validation}
\end{figure}

This paper uses the general concepts of PKIs and trust lists to maintain a trust environment for equipment compliance. The basis is the ability to validate  digital signature with the help of a certificate as depicted in Fig.~\ref{fig:cert-validation}.

Compliance of WLAN equipment has been investigated by Europan market surveillance in \cite{adcortte2013} and \cite{adcored2019}. The most prominent issue was found to be the detection of radars through the dynamic frequency selection (DFS) requirement. WiFi Alliance published a best practice document on this issue\cite{wifi-dfs-best-practise}. Technical issues of radar detection, especially when in transmission, are studied in \cite{safavi-naeini2015}. These market studies and technical studies serve only as background and motivation for this paper. In the following we will not elaborate on the technical issues, but only distinguish between ``compliant'' and ``non-compliant'' devices.

\section{Concept and Overall Architecture} \label{sec:concept}

\begin{figure}[htbp]
\centerline{\includegraphics[width=\linewidth]{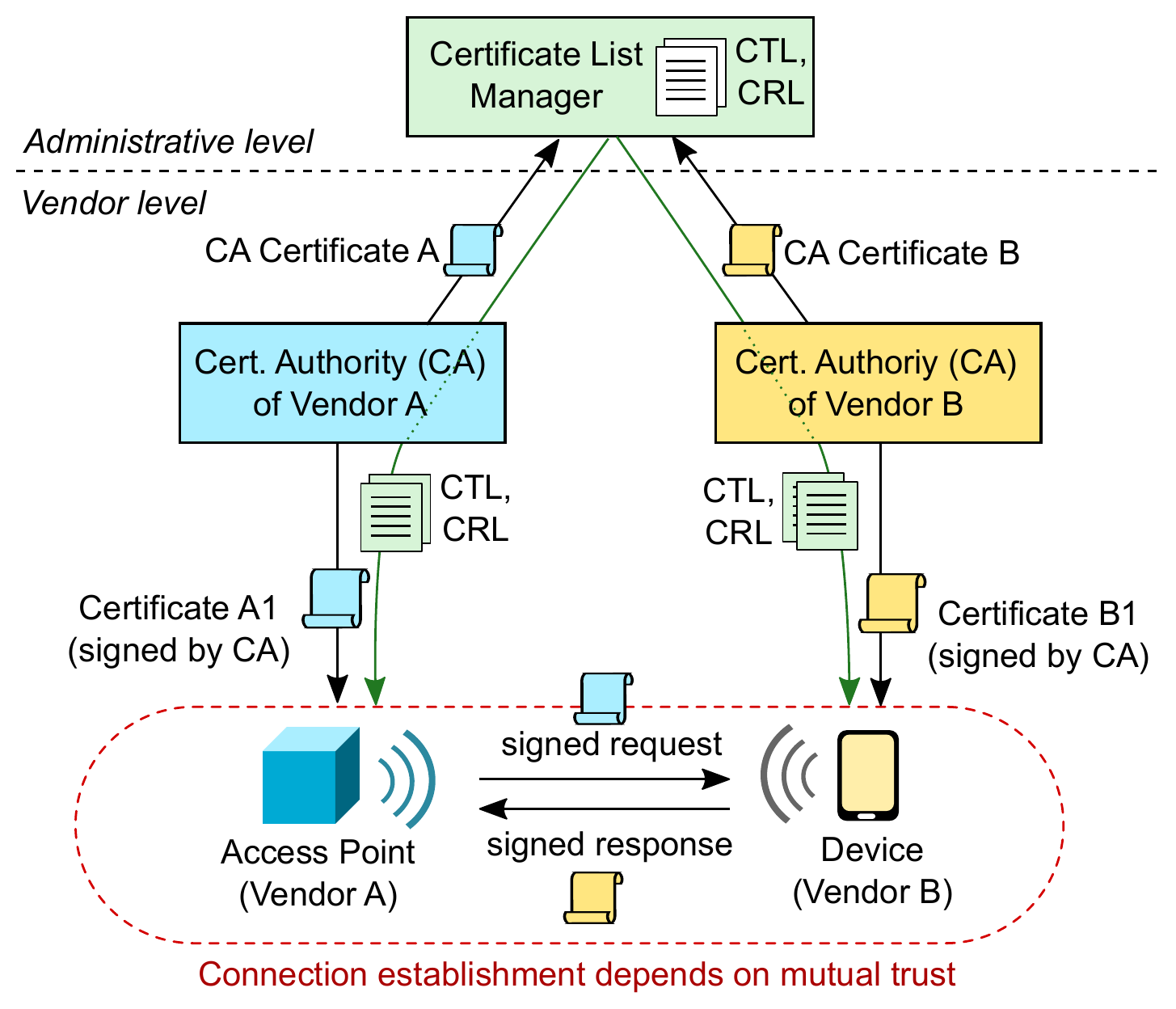}}
\caption{Overall concept of the public key infrastructure}
\label{fig:PKI}
\end{figure}

The overall architecture consists of a public key infrastructure with a Certification Authority (CA) maintained by a manufacturer or vendor (or a group of vendors). This CA issues certificates, i.e. it signs public keys of individual devices. These digital certificates are stored on each individual devices. A device by another manufacturer can verify the certificate if it knows the public key of the issuing certification authority (of the other vendor). Though each device has an individual certificate, it is not necessary to know all other device certificates, but only the higher level CA certificate that is necessary to validate a digital signature. That is achieved by collecting all CA certificates in a certificate trust list (CTL), that is distributed regularly to all devices as depicted in Figure~\ref{fig:PKI}. Certificates of non-compliant devices are put on a certificate revocation list (CRL). Regular distribution of CTL and CRL to all devices ensure that only compliant devices participate in the trust environment. A regular update does not mean a daily update, but a longer time duration that depends on the frequency of CTL changes (i.e. when new device models are registered) and CRL changes (i.e. when non-compliance cases are found).

The necessary steps are as follows:
\begin{enumerate}
\item Digital certificates are issued by a Certification Authority (CA) and stored in each device at manufacturing time.
\item All CA certificates across all vendors are collected in the certificate trust list (CTL). This trust list is provided to RLAN devices, e.g. as part of regular software updates.
\item Upon connection establishment between two RLAN devices, one device is able to verify the certificate of the other device. This could be performed in a request-response fashion after association or part of the network authentication. It could also be separately queried via protocols and data elements for device monitoring. If a valid signature and thus a valid certificate is found, then the 
\item If a certain model of RLAN equipment is found to be non-compliant by market surveillance authority, then its certificate is revoked and added to the revocation list (CTL). Once a devices obtains the updated CTL, it will reject the non-compliant device. As a consequence, the non-compliant model is limited in operation. 
\end{enumerate}

\subsection{Administration of CRL and CTL}

CTL and CRL are managed on an administrative level, which is independent of individual vendors. This could be an independent organisation, a (govenment) agency or an industry association. CTL and CRL are public, they do not need to be encrypted, but digitally signed. Thus, market surveillance authorities can check whether non-compliance cases are correctly handled. 

Certificates in the CTL and CRL can be organised on two levels, on manufacturer level and on device type level:

\textit{Option 1}: Each manufacturer maintains only one root certificate. The certificate trust list (CTL) holds one certificate for each manufacturer. In this case the CTL is only updated when a new manufacturer is registered, but on the level of the CTL a certificate revocation affects all devices of a manufacturer.

\textit{Option 2}: Each manufacturer maintains a root certificate and derived certificates for each device type or model. The certificate trust list (CTL) holds one certificate per device type or model. In this case the CTL is larger and needs more frequent updates when new equipment is brought to the market. However, on the level of the CTL, the certificate of a single non-compliant device type or model can be revoked without affecting other equipment of the same manufacturer. 

Since the device itself is subject to conformance, option 2 might be a preferred option. 

\begin{figure}[htbp]
\centerline{\includegraphics[width=\linewidth]{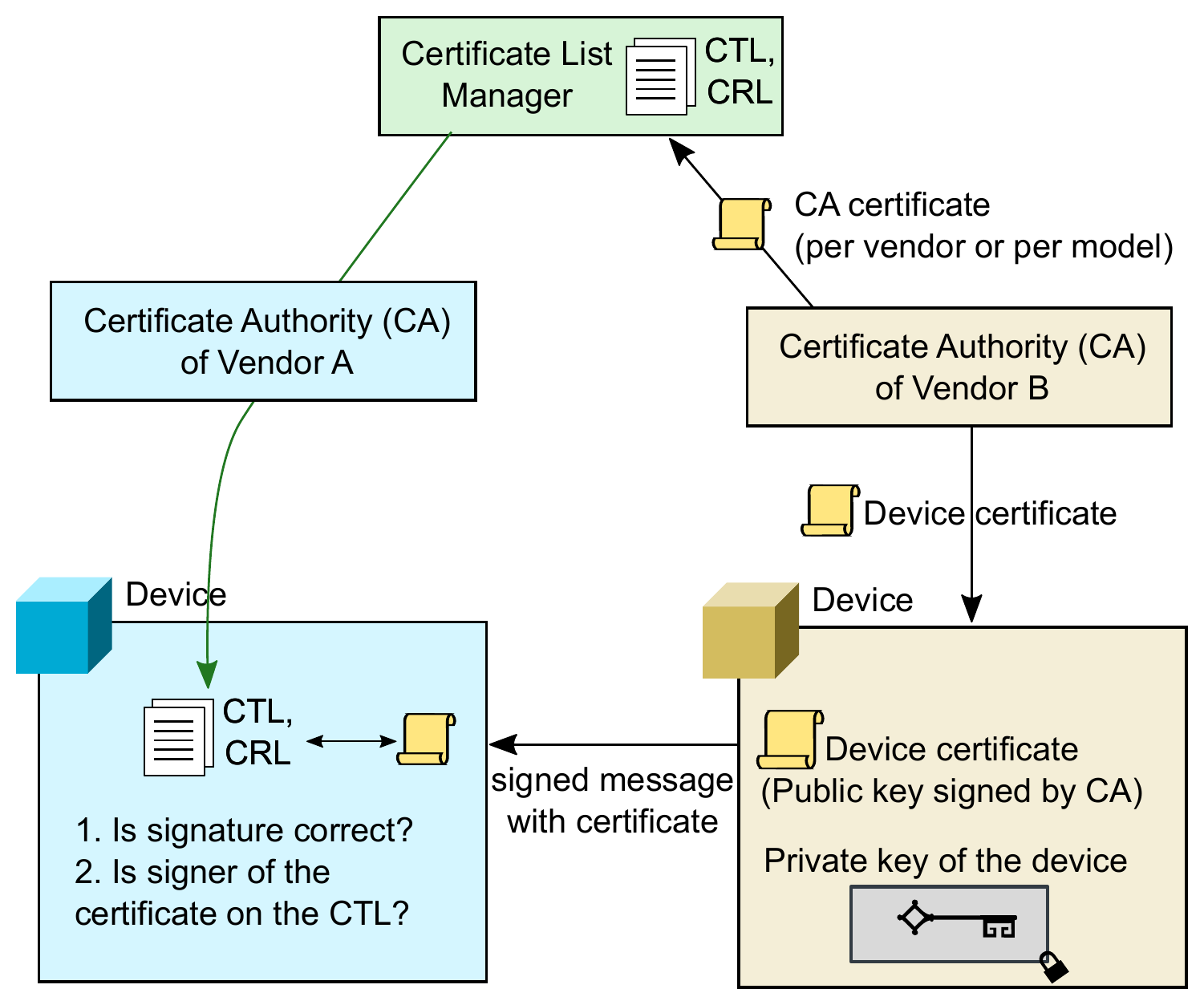}}
\caption{Validation of certificates: The receiver of a signed message with a certificate 
validates the signature and the certificate chain via the CTL.}
\label{fig:keys}
\end{figure}

\subsection{Vendor Level}

On the vendor level the certificate distribution and the device's mechanism for certificate exchange are implemented.

\begin{figure}[htbp]
\centerline{\includegraphics[width=\linewidth]{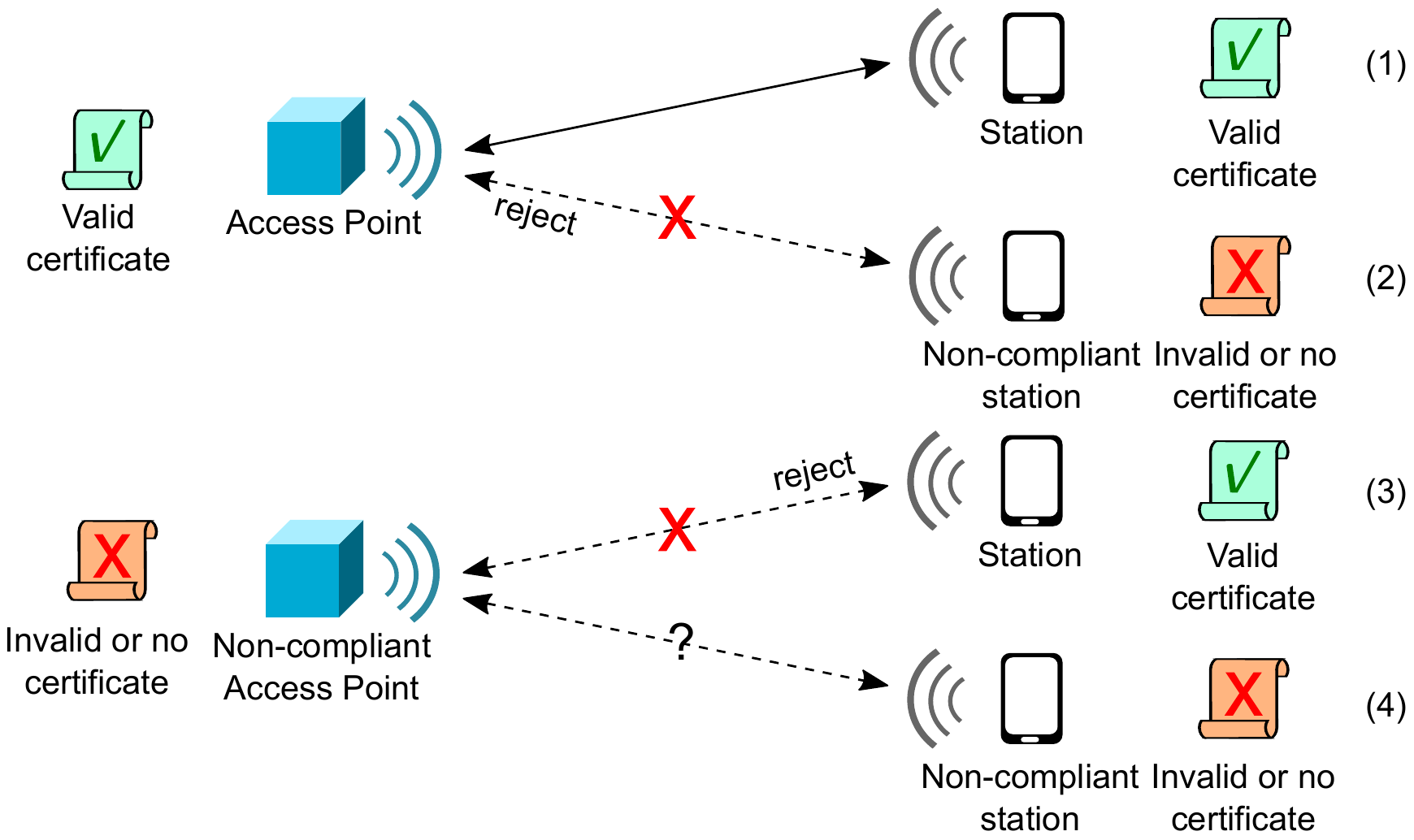}}
\caption{Connections are rejected by the compliant device if the counterpart does not hold a valid certificate. If both devices are non-compliant, no result can be stated.}
\label{fig:rejection}
\end{figure}

Connections of compliant devices to other wireless devices should only b established, when the certificate of the counterpart is valid as depicted in Fig.~\ref{fig:rejection}, with the aim to isolate non-compliant devices. Non-compliant devices, however, do not follow the rules by definition. Therefore, two non-compliant devices may or may not communicate with each other. 

The majority of radio local area networks consist of access points and stations. Usually,   station authenticates itself to the access point, i.e. it is performed in one direction. For the overall concept in this paper, however, it is beneficial that both devices check the certificate of their counterparts, i.e. certificates are validated in both directions.

Thus, devices need to have the following capabilities: 
\begin{itemize}
\item Store private key and (public) certificate as well as CTL and CRL
\item Produce digital signatures using the own private key and present signature and certificate to another querying device.
\item Validate a signature and a certificate from another device and verify the certificate chain using CTL and CRL
\item Securely update CTL and CRL
\end{itemize}
These capabilities are common for systems participating in a PKI. The exchange of digital signature and certificate might be part of the connection establishment. However, in order to avoid modification of low level authentication, it would be beneficial to use a separate protocol and check the certificate validity after the devices are associated.

\section{Discussion}  \label{sec:discussion}

While public key infrastructures are already used in many application areas, the use of a PKI for an environment of compliance has specific aspects which we discuss in the following:
\begin{itemize}
\item The PKI operation is independent of the product certification process. Vendors can bring equipment to the market, and can provide their equipment with valid certificates. 
\item Non-compliant models, if identified by market surveillance, can be blacklisted through the certificate revocation list (CRL). Then all devices of a certain model (type) can be blocked from obtaining new certificates.
\item The CTL is essential for certificate validation. Its size depends on which certificates are stored in it: manufacturer's certificates, device type certificates, or individual device certificates. While manufacturer's certificate would not allow isolation of certain non-compliant models, it is unrealistic to maintain millions of device certificates. Device types or models seem a reasonable choice.
\item Blacklisting does not immediately turn off a non-compliant device. It might still operate, but other equipment will restrict its communication. 
\item If digital certificates on the CTL and CRL are bound to device types or models (and not to manufacturers), then blacklisting affects all devices of this model. A single device could only be blacklisted through certificate revocation by the manufacturer's certification authority. 
\item A simple re-registration of a non-compliant device under a new model number needs to be prevented. Otherwise, an effective blacklisting can be circumvented.
\end{itemize}
The concept does not require any specific addition to existing public key infrastructures. However, it needs to be implemented on all equipment, including compliant equipment. Therefore it is beneficial to re-use existing protocols and data element standards that are implemented on many devices. As an example, many devices are able to establish TLS connections and the handling of X.509 certificates is part of their implementation. 

\section{Conclusion} \label{sec:conclusion}

This paper describes a concept and method for using public key infrastructures to establish a trust environment for wireless devices with respect to compliance to radio regulation. The concept distinguishes between a vendor level and and administrative level. On the administrative level, non-compliant communication equipment can be blacklisted through a certificate revocation list. On the vendor level, the distribution of certificates and the device's rejection of communication links to non-compliant equipment without valid certificate is implemented. The aim of rejecting wireless communication to non-compliant equipment is the isolation or limitation of using non-compliant air interfaces.
The proposed public key infrastructure does not enforce compliance, but it can be regarded as a tool to manage an environment of compliant wireless devices. While non-compliance needs to be detected by the responsible authority, the PKI helps to isolate non-compliant equipment, with the aim of limiting its use.

\end{document}